\documentclass[aps,pra,twocolumn,showpacs,amsmath,amssymb,superscriptaddress,notitlepage]{revtex4-1}

\usepackage{graphicx}
\usepackage{dcolumn}
\usepackage{bm}
\usepackage{comment}
\usepackage{amsmath}

\begin{document}

\title{String order via Floquet interactions in atomic systems}
\date{\today}

\author{Tony E. Lee}
\affiliation{Department of Physics, Indiana University Purdue University Indianapolis (IUPUI), Indianapolis, Indiana 46202, USA}
\author{Yogesh N. Joglekar}
\affiliation{Department of Physics, Indiana University Purdue University Indianapolis (IUPUI), Indianapolis, Indiana 46202, USA}
\author{Philip Richerme}
\affiliation{Department of Physics, Indiana University, Bloomington, Indiana 47405, USA}

\begin{abstract}
We study the transverse-field Ising model with interactions that are modulated in time. In a rotating frame, the system is described by a time-independent Hamiltonian with many-body interactions, similar to the cluster Hamiltonians of measurement-based quantum computing. In one dimension, there is a three-body interaction, which leads to string order instead of conventional magnetic order. We show that the string order is robust to power-law interactions that decay with the cube of distance. In two and three dimensions, there are five- and seven-body interactions. We discuss adiabatic preparation of the ground state as well as experimental implementation with trapped ions, Rydberg atoms, and polar molecules.
\end{abstract}

\maketitle

\section{Introduction}

A current goal in atomic physics is to realize exotic many-body phases, since atomic-physics experiments can simulate the physics of condensed-matter systems \cite{pachos04,bermudez09,gorshkov11,lee13,yan13,richerme14,cohen14,daley14,chan15,schauss15,gong16,kaczmarczyk16}. An advantage of atomic quantum simulators is that they are highly tunable. For example, the interaction between atoms is often induced by a laser, so one can easily tune the interaction strength, sign, and range by changing the laser parameters \cite{molmer99,porras04}.

An intriguing type of many-body phase is symmetry-protected topological order \cite{chen11}. A common feature of such a phase is string order. A system with string order does not appear to have long-range order according to two-site correlation functions. However, if one calculates a nonlocal correlation function involving a long ``string'' of operators, the hidden order becomes apparent. A well-known example of string order is the Haldane phase of a spin-1 chain \cite{haldane83,dennijs89,kennedy92,else13}.

In this paper, we show that one can realize string order with spin-1/2 particles by modulating the interaction of the transverse-field Ising model. This scheme is well-suited for atomic-physics experiments, since it exploits their tunability. In one-dimension, a spin chain with time-modulated, two-body interactions is equivalent to a spin chain with time-independent, three-body interactions. The three-body interaction leads to string order in the ground state. The three-body interaction is a cluster Hamiltonian of measurement-based quantum computing, and the string order is a manifestation of the cluster phase \cite{raussendorf01,nielsen06}. We discuss how to adiabatically prepare the ground state in the laboratory frame.

We then show that the scheme works even if the original spin chain has power-law interactions, as is common in atomic-physics experiments \cite{yan13,richerme14,schauss15,gong16}. For interactions that decay with the cube of distance, the long-range interactions have negligible effect on the ground state, and the ground state still has string order. We also show that modulating the interaction in two and three dimensions leads to five- and seven-body interactions. Finally, we discuss experimental implementation with trapped ions, Rydberg atoms, and polar molecules.

In recent years, there has been a lot of work on using time-periodic modulation to control many-body systems (see Refs.~\cite{bukov15,eckardt16} for recent reviews). The generation of many-body interactions has been previously studied in the context of bosonic quantum gases \cite{gong09,rapp12,greschner14,meinert16} and spin models \cite{engelhardt13,iadecola15}. Other time-modulated spin models have also been studied \cite{bastidas12,russomanno15,gomez15}.

\section{General model} \label{sec:general}

We consider a lattice of spin-1/2 particles, where the spin-spin interaction is modulated in time. The Hamiltonian is ($\hbar=1$)
\begin{eqnarray}
H&=&\frac{J\cos(\Omega t)}{2} \sum_{mn} a_{mn} X_m X_n + g \sum_n Z_n, \label{eq:H}
\end{eqnarray}
where $X_n,Y_n,Z_n$ are the Pauli matrices for spin $n$, $J$ is the modulation amplitude, and $g$ is the transverse field strength. The 1/2 accounts for double-counting pairs in the sum. For now, we let the lattice be arbitrary, and $a_{mn}$ encodes the connectivity between spins $m$ and $n$. In later sections, we will consider one, two, and three-dimensional lattices. Reference \cite{engelhardt13} studied the case of all-to-all coupling. Reference \cite{iadecola15} studied the case of one dimension with nearest-neighbor interactions.

For convenience, we define the operator ${A_n\equiv\sum_m a_{mn} X_m}$, which is the sum of the spins that spin $n$ interacts with. Then Eq.~\eqref{eq:H} can be written as
\begin{eqnarray}
H&=&\frac{J\cos(\Omega t)}{2} \sum_{n} A_n X_n + g \sum_n Z_n. \label{eq:H_Fn}
\end{eqnarray}

To analyze this time-dependent model, we use Floquet theory \cite{bukov15,eckardt16}. We go into the interaction picture, rotating with the first term of Eq.~\eqref{eq:H_Fn}. In this rotating frame, $H$ becomes $H'$:
\begin{eqnarray}
H'&=&U(t)^\dagger \left(g \sum_n Z_n \right) U(t),\\
U(t)&=&\exp\left[-\frac{i\beta(t)}{2} \sum_n A_n X_n\right], \label{eq:U}\\
\beta(t) &=& \frac{J}{\Omega}\sin(\Omega t). \label{eq:a}
\end{eqnarray}
Then we use the Baker-Campbell-Hausdorff formula to rewrite $H'$,
\begin{eqnarray}
H'&=&\sum_n g[\cos(2\beta(t) A_n)Z_n + \sin(2\beta(t) A_n)Y_n],\\
&=&\sum_n \frac{g}{2}[(Z_n-iY_n) e^{i2\beta(t) A_n} + (Z_n+iY_n) e^{-i2\beta(t) A_n}].\nonumber\\
\end{eqnarray}
We write in terms of Bessel functions:
\begin{eqnarray}
\exp\left[\frac{2iJ}{\Omega}\sin(\Omega t)A_n\right]&=& \sum_{\ell=-\infty}^\infty \mathcal{J}_\ell\left(\frac{2JA_n}{\Omega}\right) e^{i\ell\Omega t}.
\end{eqnarray}
$H'$ is still time dependent, so we make a rotating-wave approximation \cite{ashhab07,bastidas12}. The $\ell\neq0$ terms in $H'$ oscillate very quickly and are off-resonant. Thus, we only need to keep the $\ell=0$ terms to capture the slow-time-scale dynamics. This rotating-wave approximation is valid when
\begin{eqnarray}
\Omega\gg g. \label{eq:rwa_condition}
\end{eqnarray}
So the final Hamiltonian is
\begin{eqnarray}
H'&=&g\sum_n \mathcal{J}_0\left(\frac{2JA_n}{\Omega}\right)Z_n,\label{eq:Hp}\\
&=&g\sum_n \mathcal{J}_0\left(\frac{2J}{\Omega}\sum_m a_{mn} X_m\right)Z_n.\label{eq:Hp2}
\end{eqnarray}

Thus, in the interaction picture and when $\Omega$ is sufficiently large, the system is described by the time-independent Hamiltonian $H'$ in Eq.~\eqref{eq:Hp2}. At this point, we choose a lattice geometry ($a_{mn}$) and expand $\mathcal{J}_0$ in a power series:
\begin{eqnarray}
\mathcal{J}_0(x)&=&\sum_{p=0}^\infty \frac{(-1)^p}{p!\,\Gamma(p+1)}\left(\frac{x}{2}\right)^{2p}.
\end{eqnarray}
We obtain arbitrary even powers of $A_n$, which can be simplified using the fact that $X_m^2=1$. In general, $H'$ has many-body interactions involving one $Z$ and an even number of $X$'s, e.g., $X_1Z_2X_3$. $H'$ is reminiscent of cluster Hamiltonians that arise in measurement-based quantum computing \cite{raussendorf01,nielsen06}.

The presence of many-body interactions in $H'$ can be intuitively understood as follows. The transverse field in $H$ causes spin $n$ to undergo Rabi oscillations between $\left|\downarrow\right\rangle_x$ and $\left|\uparrow\right\rangle_x$. However, the modulated interaction means that spin $n$ sees an oscillating energy shift that depends on its neighbors' $X$. Similarly, the many-body terms in $H'$ mean that spin $n$ Rabi-oscillates depending on its neighbors' $X$, but now the energy shift is time-independent.

The relationship between the wave function $|\psi\rangle$ in the laboratory frame (evolving with $H$) and the wave function $|\psi'\rangle$ in the rotating frame (evolving with $H'$) is:
\begin{eqnarray}
|\psi(t)\rangle&=&U(t)|\psi'(t)\rangle.
\end{eqnarray}
Equations \eqref{eq:U} and \eqref{eq:a} say that when $t$ is a multiple of $2\pi/\Omega$, $\beta(t)=0$ and $U(t)=1$. Thus, if we measure the system at these periodic times, $|\psi\rangle=|\psi'\rangle$ and we do not have to worry about converting between the two frames \cite{engelhardt13}.

There is a simple way to convert between the two frames at arbitrary times. We note that $|\psi'(t)\rangle=U(t)^\dagger|\psi(t)\rangle$, where $U(t)^\dagger$ is the evolution operator of $H$ with $g=0$ and $J\rightarrow -J$. Thus, after we have obtained $|\psi(t)\rangle$, if we evolve it further for time $t$ with $g=0$ and $J\rightarrow -J$, we obtain $|\psi'(t)\rangle$.

In Eq.~\eqref{eq:H}, we assumed that the interaction alternates sign, but our results still hold if the interaction is modulated without changing sign. In that case, $H'$ is the same, but $U(t)$ is different \cite{bastidas12}. In some experimental setups, it is easier to modulate the strength without changing the sign.

Lastly, we note that although we assume spin-1/2 in this paper, one obtains similar results for higher spin. Suppose the $X_n,Y_n,Z_n$ in Eq.~\eqref{eq:H} were for higher spin. Then $H'$ would still be given by Eq.~\eqref{eq:Hp2}, since the commutation relations of $X_n,Y_n,Z_n$ do not depend on the spin magnitude. The difference is that $X_n^2\neq 1$ for higher spin, so the expanded and simplified form of $H'$ would look different.

\section{One dimension with nearest-neighbor interactions}

\subsection{Model}

We now consider a one-dimensional lattice of $N$ spins with nearest-neighbor interactions, which was first studied in Ref.~\cite{iadecola15}. We assume open boundary conditions. However, we add a longitudinal field to the edge spins:
\begin{eqnarray}
H&=&J\cos(\Omega t)\left[ \sum_{n=1}^{N-1} X_n X_{n+1} + X_1 + X_N \right]+ g \sum_{n=1}^{N} Z_n. \nonumber\\ \label{eq:H_1d}
\end{eqnarray}
It is not necessary to add $X_1+X_N$, but without these extra terms, the ground state of $H'$ is fourfold degenerate due to a $\mathbb{Z}_2\times\mathbb{Z}_2$ symmetry \cite{iadecola15,son11,smacchia11}. Although this degeneracy is a signature of symmetry-protected-topological order, it is problematic if one wants to adiabatically prepare the ground state of $H'$; without these terms, one needs a slower ramp. Another reason for adding these terms is to make $H'$ look more like a cluster Hamiltonian, as discussed below. 

\begin{figure}[t]
\centering
\includegraphics[width=3.7 in,trim=1.8in 4.1in 1.6in 4.2in,clip]{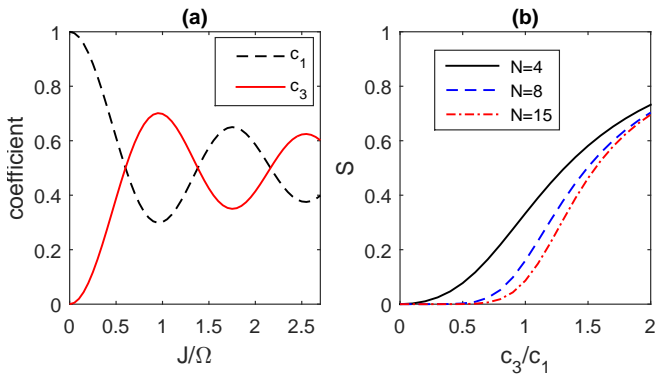}
\caption{\label{fig:1d_nn}One-dimensional chain with nearest-neighbor interactions. (a) Coefficients in $H'$, in units of $g$. (b) String-order parameter for ground state of $H'$.}
\end{figure}

The non-edge spins have $A_n=X_{n-1}+X_{n+1}$, while the edge spins have $A_1=1+X_2$ and $A_N=X_{N-1}+1$. After expanding and simplifying Eq.~\eqref{eq:Hp2}, we obtain \cite{iadecola15}
\begin{eqnarray}
H'&=&c_1 \sum_{n=1}^N Z_n - c_3\Bigg(\sum_{n=2}^{N-1} X_{n-1}Z_n X_{n+1} \nonumber\\
&&\quad\quad\quad\quad\quad\quad\quad+ Z_1X_2 + X_{N-1}Z_N\Bigg), \label{eq:Hp_1d}\\
c_1 &=& \frac{g}{2}\left[1+\mathcal{J}_0\left(\frac{4J}{\Omega}\right)\right], \label{eq:c1_1d}\\
c_3 &=& \frac{g}{2}\left[1-\mathcal{J}_0\left(\frac{4J}{\Omega}\right)\right]. \label{eq:c3_1d}
\end{eqnarray}
Thus, $H'$ has a transverse field with strength $c_1$ and a three-body interaction with strength $c_3$. The ratio $c_3/c_1$ can be adjusted by varying $J/\Omega$ [Fig.~\ref{fig:1d_nn}(a)]. The transverse field is always present, although it can be weaker than the three-body interaction. $c_3/c_1$ reaches its maximum value of 2.35 when $J/\Omega=0.96$.

\subsection{Cluster phase}

$H'$ in Eq.~\eqref{eq:Hp_1d} is significant for two reasons. First, the terms in parentheses are a cluster Hamiltonian \cite{pachos04,doherty09,skrovseth09,son11,smacchia11,giampaolo15}. The (unique) ground state of a cluster Hamiltonian is a cluster state, which is a highly entangled state that is useful for measurement-based quantum computing \cite{raussendorf01,nielsen06}. Note that in order to be a cluster Hamiltonian, $H'$ must have boundary terms as in Eq.~\eqref{eq:Hp_1d}, which is why we included $X_1+X_N$ in Eq.~\eqref{eq:H_1d}.

$H'$ is also significant because the ground state exhibits a phase transition to string order. The string-order parameter is \cite{son11,smacchia11,giampaolo15}
\begin{eqnarray}
S=\lim_{N\rightarrow\infty}\left|\left\langle X_1Y_2 \left(\prod_{k=3}^{N-2} Z_k\right) Y_{N-1}X_N\right\rangle\right|, \label{eq:S}
\end{eqnarray}
although other string-order parameters also work \cite{doherty09,skrovseth09}. When $c_3/c_1>1$, the ground state has string order ($S>0$), and the system is in the ``cluster phase,'' since the ground state is still useful for measurement-based quantum computing even if $c_1\neq0$ \cite{doherty09}. When $c_3/c_1<1$, the ground state is in the paramagnetic phase ($S=0$). The critical point at $c_3/c_1=1$ is a second-order phase transition. These properties can be obtained analytically by using the Jordan-Wigner transformation \cite{pachos04,smacchia11} or by mapping to the (time-independent) transverse-field Ising model \cite{doherty09,son11}. Note that two-site correlations, such as $\langle X_mX_n\rangle$, do not show long-range order \cite{pachos04}.

There is in fact a deep connection between string order and measurement-based quantum computing \cite{doherty09}. In measurement-based quantum computing, one does a sequence of local measurements to entangle distant qubits. The sequence of local measurements is equivalent to a string operator. If a state's string-order parameter is nonzero, the state is useful for measurement-based quantum computing because local measurements on it produce a state that is more entangled than a random state. A larger string-order parameter implies more usefulness.

Figure \ref{fig:1d_nn}(b) shows the string-order parameter for finite $N$, calculated using exact diagonalization of $H'$. As $N$ increases, the phase transition at $c_3/c_1=1$ becomes more evident.



\subsection{Adiabatic preparation} \label{sec:adiabatic}

\begin{figure}[b]
\centering
\includegraphics[width=4 in,trim=1.6in 4.3in 1in 4.3in,clip]{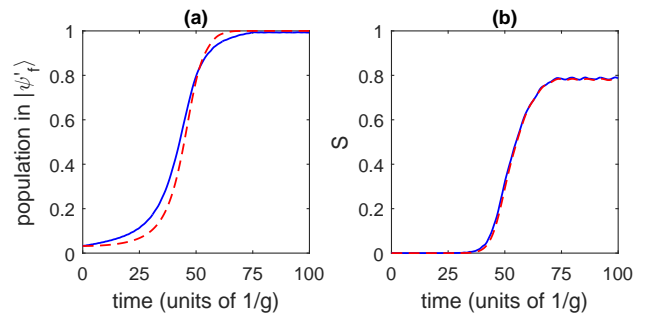}
\caption{\label{fig:adiabatic_nn}Adiabatic ramp from $J=0$ to $0.96\Omega$ in the laboratory frame (blue solid line) and rotating frame (red dashed line). (a) Population in the target final state $|\psi'_f\rangle$. (b) String-order parameter. In the laboratory frame, the wave function is sampled stroboscopically in time at multiples of $2\pi/\Omega$. We use a 1D chain with $N=8$, nearest-neighbor interactions, $t_\text{ramp}=75/g$, and $\Omega=10g$. }
\end{figure}

Now we discuss how to adiabatically prepare the cluster phase of $H'$ by turning on the three-body interaction. We linearly increase $J$ from 0 to $0.96\Omega$ over a time $t_\text{ramp}$, such that $c_3/c_1$ starts at 0 and ends at the maximum value of 2.35. The phase transition at $c_3/c_1=1$ corresponds to $J=0.60\Omega$. The system starts in $|\psi\rangle=|\psi'\rangle=\left|\downarrow\downarrow\downarrow\cdots\right\rangle$, which is the initial paramagnetic ground state of $H'$. We denote the final ground state of $H'$ as $|\psi'_f\rangle$, which is the desired cluster phase.

We first simulate the adiabatic ramp in the rotating frame ($|\psi'\rangle$ evolves with $H'$). Figure \ref{fig:adiabatic_nn} shows there is a clean transfer of population into $|\psi'_f\rangle$, as expected. There is a slight infidelity, which can be reduced by using a slower ramp (larger $t_\text{ramp}$).

Next, we simulate the adiabatic ramp in the laboratory frame ($|\psi\rangle$ evolves with $H$). We sample $|\psi\rangle$ at periodic times, so that the wave functions of the laboratory and rotating frames coincide. Figure \ref{fig:adiabatic_nn}(a) shows that the population transfer in the laboratory frame is similar to the rotating frame. The deviation is due to the rotating-wave approximation, but the agreement improves as $\Omega$ increases. Interestingly, despite the deviation from $|\psi'\rangle$ during the ramp, $|\psi\rangle$ ends up in $|\psi'_f\rangle$ with very high fidelity.

\begin{figure*}[t]
\centering
\includegraphics[width=6.9 in,trim=0.5in 2.6in 0.5in 2.6in,clip]{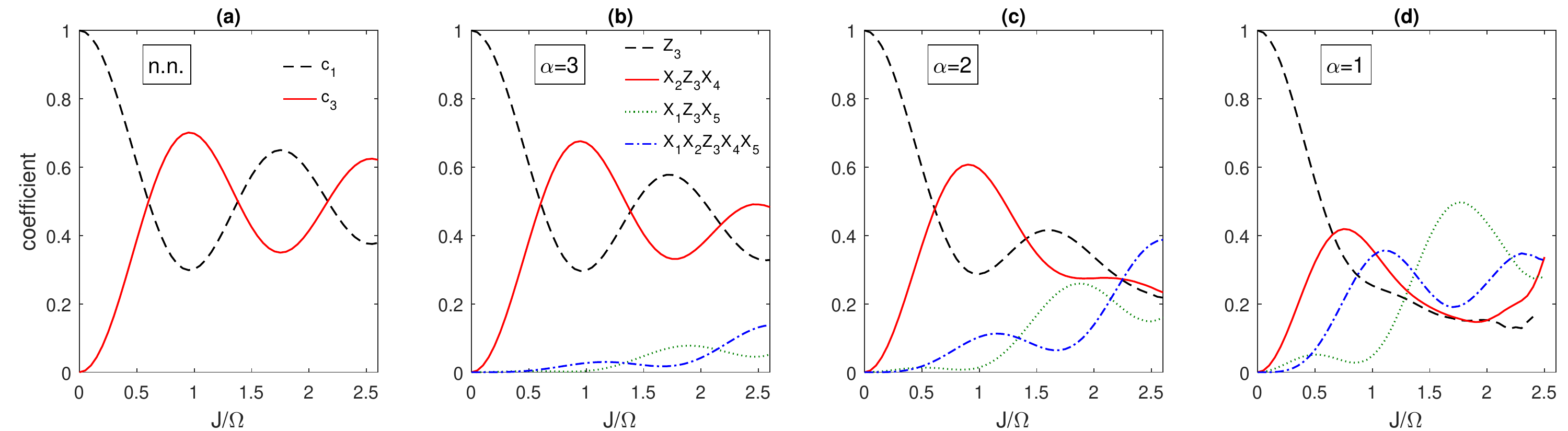}
\caption{\label{fig:coeff_power_law}Coefficients in $H'$ for a 1D chain of $N=5$ spins with different interaction ranges: (a) nearest neighbor, (b) power law with $\alpha=3$, (c) power law with $\alpha=2$, and (d) power law with $\alpha=1$. Coefficients are in units of $g$.}
\end{figure*}

In Fig.~\ref{fig:adiabatic_nn}(b) we plot the string-order parameter $S$ as a function of time. As expected, $S$ starts at zero and ends at a nonzero value.

It turns out that this adiabatic method does not work for $N=5,9,13,\ldots$, due to a symmetry of $H'$. If one starts the adiabatic ramp in $\left|\downarrow\downarrow\downarrow\cdots\right\rangle$, the system ends up in an excited state.

\section{One dimension with power-law interactions} \label{sec:1d_power}

We now consider a one-dimensional lattice with power-law interactions, i.e., the spin-spin interaction decreases with a power law in distance ($1/r^\alpha$). The motivation is that atom-based quantum simulators (e.g., trapped ions \cite{richerme14}, Rydberg atoms \cite{schauss15}, and polar molecules \cite{gorshkov11}) usually have power-law interactions. Below, we present results for $\alpha=1,2,3$, which are relevant to these experiments. We show that $\alpha=3$ is almost identical to the nearest-neighbor model, whereas $\alpha=1$ is quite different.

\subsection{Model}

We assume open boundary conditions and again add a longitudinal field to the edge spins:
\begin{eqnarray}
H&=&J\cos(\Omega t)\left[ \sum_{\substack{m,n=1\\m<n}}^{N} \frac{1}{|m-n|^\alpha}X_m X_n + X_1 + X_N \right] \nonumber\\ 
&&+ g \sum_{n=1}^{N} Z_n.\label{eq:H_1d_power}
\end{eqnarray}
So for non-edge spins, $A_n=\sum_{m\neq n} X_m/|m-n|^\alpha$.

We now calculate $H'$ for Eq.~\eqref{eq:H_1d_power}. Since $H$ now has coupling between every pair of spins, $H'$ has many more terms than Eq.~\eqref{eq:Hp_1d}. There are three-body terms between non-neighboring spins. There are also many-body interactions involving five, seven, etc., spins. There are two questions we seek to answer. First, how large are these extra terms? Second, how does the ground state of the new $H'$ compare to that of the nearest-neighbor case [Eq.~\eqref{eq:Hp_1d}]? 

To get a sense of the magnitude of the extra terms, we consider in detail the case of $N=5$ spins, which is representative of larger $N$. In this case, $H'$ has one-, three-, and five-body terms. Figure \ref{fig:coeff_power_law} plots the coefficients for some terms that involve $Z_3$. We calculated these coefficients numerically by expanding Eq.~\eqref{eq:Hp2} to sufficient order. We see that the extra terms are small for $\alpha=3$ but large for $\alpha=1$. So $H'$ for $\alpha=3$ is very similar to the nearest neighbor case, whereas $\alpha=1$ is quite different.

We discuss in detail the case of $\alpha=3$. For $J/\Omega\lesssim2$, the coefficients of $Z_3$ and $X_2Z_3X_4$ are very close to Eqs.~\eqref{eq:c1_1d} and \eqref{eq:c3_1d}, whereas the coefficients of the extra terms are small. If we set $J/\Omega=0.96$ (which maximized $c_3/c_1$ for the nearest-neighbor case), the extra terms are very small. The largest extra term is $X_1X_2Z_3X_4X_5$, whose coefficient is still 1/30 that of $X_2Z_3X_4$. Also, the new three-body terms are smaller than what one would naively expect based on the cubic power law. For example, one would expect the coefficient of $X_1 Z_3 X_5$ to be $1/64$ that of $X_2 Z_3 X_4$ (since the $X_1X_3$ and $X_3X_5$ interactions in $H$ are 1/8 those of $X_2X_3$ and $X_3X_4$), but the ratio is actually $1/240$. So the power-law decay of interaction in $H$ does not directly carry over to $H'$.

Although Fig.~\ref{fig:coeff_power_law} only shows terms involving $Z_3$, we observe similar behavior for other $Z_n$. Furthermore, as we increase $N$, the above observations still hold. So $H'$ for $\alpha=3$ is very close to that for the nearest neighbor case [Eq.~\eqref{eq:Hp_1d}]. However, it is possible that the ground states are very different, so we proceed to compare the ground states.

\subsection{Ground state of $H'$}

Here, we compare the ground state $|\psi'_{pl}\rangle$ of $H'$ for the power-law case with the ground state $|\psi'_{nn}\rangle$ of $H'$ for the nearest-neighbor case [Eq.~\eqref{eq:Hp_1d}]. In principle, we could find $|\psi'_{pl}\rangle$ by first calculating all terms of $H'$ via Eq.~\eqref{eq:Hp2}, then diagonalizing to find the ground state, but this is very tedious for large $N$. A more convenient way is to perform an adiabatic ramp of $H$, such as in Sec.~\ref{sec:adiabatic}: If the ramp is very slow  (to ensure adiabaticity) and $\Omega$ is very large (to validate the rotating-wave approximation), then $|\psi'_{pl}\rangle$ will be prepared with very high fidelity. (At the moment, we are interested in the ideal case in order to obtain the ground state of $H'$. In Sec.~\ref{sec:numbers}, we will use more realistic experimental parameters.)  Note that, as discussed in Sec.~\ref{sec:adiabatic}, the ground states for $N=5,9,13,\dots$, are inaccessible via adiabatic ramp.

\begin{figure}[t]
\centering
\includegraphics[width=4 in,trim=1.6in 4.3in 1in 4.3in,clip]{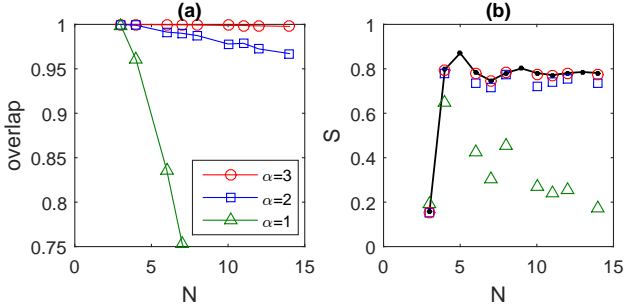}
\caption{\label{fig:comparison_nn_pl} Comparison of ground states of $H'$ with different interaction power laws for $J/\Omega=0.96$. (a) Overlap of ground states with nearest-neighbor ground state, $|\langle\psi'_{nn}|\psi'_{pl}\rangle|^2$. (b) String-order parameter for nearest-neighbor (black points) and power-law interactions.}
\end{figure}

Figure \ref{fig:comparison_nn_pl} compares $|\psi'_{pl}\rangle$ with $|\psi'_{nn}\rangle$ for $J/\Omega=0.96$. In these plots, $|\psi'_{pl}\rangle$ was obtained using $t_\text{ramp}=300/g$ and $\Omega=100g$. Figure \ref{fig:comparison_nn_pl}(a) shows the overlap $|\langle\psi'_{nn}|\psi'_{pl}\rangle|^2$ for different $N$ and $\alpha$. The case of $\alpha=3$ has very high overlap (0.998 for $N=14$), whereas the case of $\alpha=1$ has small overlap. Actually, the overlap is larger than what is plotted due to the way we obtained $|\psi'_{pl}\rangle$, i.e., using a slower ramp and larger $\Omega$ would make the overlap even higher. Figure \ref{fig:comparison_nn_pl}(b) plots the string-order parameter $S$: For $\alpha=3$, there is very good agreement between $|\psi'_{pl}\rangle$ and $|\psi'_{nn}\rangle$. Thus for $\alpha=3$,  $|\psi'_{pl}\rangle$ is almost identical to $|\psi'_{nn}\rangle$. This means that quantum simulators with cubic power-law interactions can observe the transition to string order.

\section{Higher dimension}

Now we consider two- and three-dimensional square lattices with nearest-neighbor interactions,
\begin{eqnarray}
H&=&J\cos(\Omega t) \sum_{\langle mn\rangle} X_m X_n + g \sum_n Z_n. \label{eq:H_2d}
\end{eqnarray}
For simplicity, we assume periodic boundary conditions.

\begin{figure}[t]
\centering
\includegraphics[width=3.7 in,trim=1.8in 4.2in 1.6in 4.2in,clip]{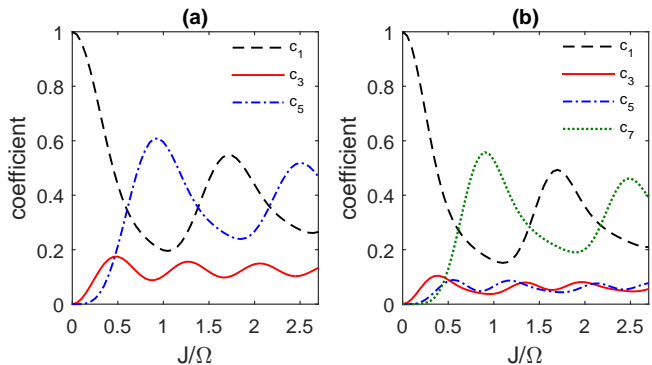}
\caption{\label{fig:coeff_2d3d} Coefficients $c_n$ of $n$-body terms in $H'$ for (a) 2D and (b) 3D square lattices. Coefficients are in units of $g$.}
\end{figure}

For a 2D lattice, expanding Eq.~\eqref{eq:Hp2} leads to
\begin{eqnarray}
H'&=&c_1 \sum_n Z_n - c_3\sum_n \sideset{}{'}\sum_{i,j\in\mathcal{N}(n)} X_i X_j Z_n  \nonumber\\
&& + c_5\sum_n  \sideset{}{'}\sum_{i,j,k,\ell\in\mathcal{N}(n)} X_i X_j X_k X_{\ell} Z_n, \label{eq:Hp_2d}\\
c_1 &=& \frac{g}{8}\left[3+4\mathcal{J}_0\left(\frac{4J}{\Omega}\right) + \mathcal{J}_0\left(\frac{8J}{\Omega}\right)\right], \label{eq:c1_2d}\\
c_3 &=& \frac{g}{8}\left[1-\mathcal{J}_0\left(\frac{8J}{\Omega}\right)\right], \label{eq:c3_2d}\\
c_5 &=& \frac{g}{8}\left[3-4\mathcal{J}_0\left(\frac{4J}{\Omega}\right) + \mathcal{J}_0\left(\frac{8J}{\Omega}\right)\right], \label{eq:c5_2d}
\end{eqnarray}
where $\mathcal{N}(n)$ denotes the nearest neighbors of spin $n$, and $\sum'$ means to include each set of neighbors only once. The $c_3$ terms are three-body interactions involving each pair of neighbors of $n$, whereas the $c_5$ terms are five-body interactions involving all four neighbors. Figure \ref{fig:coeff_2d3d}(a) plots the coefficients. There are ranges of $J/\Omega$ where $c_5>c_1,c_3$.

Equation \eqref{eq:Hp_2d} was previously studied for the case of $c_3=0$ \cite{doherty09}. There is a phase transition at $c_5/c_1=1$ with string order for $c_5/c_1>1$, where the string runs diagonally across the lattice. Note from Fig.~\ref{fig:coeff_2d3d}(a) that $H'$ always has $c_3>0$; it will be interesting to see how $c_3$ affects the phase transition described in Ref.~\cite{doherty09}.

For a 3D lattice, expanding Eq.~\eqref{eq:Hp2} leads to
\begin{eqnarray}
H'&=&c_1 \sum_n Z_n - c_3\sum_n  \sideset{}{'}\sum_{i,j\in\mathcal{N}(n)} X_i X_j Z_n  \nonumber\\
&& + c_5\sum_n  \sideset{}{'}\sum_{i,j,k,\ell\in\mathcal{N}(n)} X_i X_j X_k X_{\ell} Z_n \nonumber\\
&& - c_7\sum_n  \sideset{}{'}\sum_{i,j,k,\ell,m,p\in\mathcal{N}(n)} X_i X_j X_k X_{\ell} X_m X_p Z_n, \quad\label{eq:Hp_3d}
\end{eqnarray}
\begin{eqnarray}
c_1 &=& \frac{g}{32}\left[10+15\mathcal{J}_0\left(\frac{4J}{\Omega}\right) + 6\mathcal{J}_0\left(\frac{8J}{\Omega}\right) + \mathcal{J}_0\left(\frac{12J}{\Omega}\right)\right], \label{eq:c1_3d}\nonumber\\ \\
c_3 &=& \frac{g}{32}\left[2+\mathcal{J}_0\left(\frac{4J}{\Omega}\right) - 2\mathcal{J}_0\left(\frac{8J}{\Omega}\right) - \mathcal{J}_0\left(\frac{12J}{\Omega}\right)\right], \label{eq:c3_3d}\nonumber\\ \\
c_5 &=& \frac{g}{32}\left[2-\mathcal{J}_0\left(\frac{4J}{\Omega}\right) - 2\mathcal{J}_0\left(\frac{8J}{\Omega}\right) + \mathcal{J}_0\left(\frac{12J}{\Omega}\right)\right], \label{eq:c5_3d}\nonumber\\ \\
c_7 &=& \frac{g}{32}\left[10-15\mathcal{J}_0\left(\frac{4J}{\Omega}\right) + 6\mathcal{J}_0\left(\frac{8J}{\Omega}\right) - \mathcal{J}_0\left(\frac{12J}{\Omega}\right)\right]. \label{eq:c7_3d}\nonumber\\ 
\end{eqnarray}
So $H'$ includes up to seven-body interactions. Figure \ref{fig:coeff_2d3d}(b) plots the coefficients. There are ranges of $J/\Omega$ where $c_7>c_1,c_3,c_5$.

It is important to note that $c_5$ in Eq.~\eqref{eq:c5_2d} and $c_7$ in Eq.~\eqref{eq:c7_3d} are the same order as $c_3$ in Eq.~\eqref{eq:c3_1d}. This is surprising since one would expect interactions involving more spins to be a lot smaller. We also note that Eqs.~\eqref{eq:Hp_2d} and \eqref{eq:Hp_3d} are cluster Hamiltonians; this is significant because a measurement-based quantum computer beats a classical computer when the cluster state is on a lattice higher than 1D \cite{nielsen06}.

It turns out that the form and coefficients of $H'$ depend only on the number of neighbors in $H$. If $H$ was on a 2D triangular lattice (six neighbors), $H'$ would still be given by Eqs.~\eqref{eq:Hp_3d}--\eqref{eq:c7_3d}. This is because $A_n$ has the same form for a 2D triangular lattice as a 3D square lattice. On the other hand, a 2D honeycomb lattice (three neighbors) would have a different form.

\section{Experimental implementation}

\subsection{Possible setups}

We now discuss three types of experiments that could modulate the interaction as in Eq.~\eqref{eq:H}. In these experiments, a spin state is encoded in the levels of an atom or molecule, and the spin-spin interaction is engineered via laser fields. Time-independent interactions have been demonstrated for all three types, and introducing a modulation is straightforward. (It is easier to modulate the interaction strength without changing the sign, but our results still hold in this case, albeit with a different $U(t)$ \cite{bastidas12}.) These experiments have power-law interactions, but Sec.~\ref{sec:1d_power} showed that $\alpha=3$ ends up being very similar to the nearest-neighbor case.

The first example is trapped ions \cite{richerme14}. In this setup, a laser induces a spin-spin interaction between ions \mbox{($\alpha=$0--3)}. The sign and magnitude of the interaction depends on the frequency and intensity of the laser \cite{molmer99,porras04}. By modulating the laser parameters, one modulates the interaction. 

The second example is Rydberg atoms \cite{schauss15}. Rydberg levels have strong polar interactions. One can generate spin-spin interactions ($\alpha=3,6$) by dressing a ground state with a Rydberg state via an off-resonant laser \cite{macri14}. The sign and magnitude of the interaction depends on which Rydberg state is used and the intensity of the dressing laser. By modulating the intensity of the dressing laser, one modulates the interaction. An alternative approach is to directly populate a Rydberg state; since the polar interaction can be tuned via a F\"orster resonance \cite{barredo15}, one can modulate the interaction by modulating electric or microwave fields.

The third example is polar molecules \cite{gorshkov11,yan13}. In this case, a spin is encoded in the rotational degree of freedom of a molecule. The molecules interact via polar interactions ($\alpha=3$), which can be tuned via electric and microwave fields. By modulating the latter, one modulates the interaction.

\subsection{Experimental numbers}\label{sec:numbers}
To maximize fidelity of the prepared ground state, $\Omega$ and $t_\text{ramp}$ should both be large to ensure validity of the rotating-wave approximation and adiabaticity, respectively. In practice, these are limited because the interaction strength $J$ cannot be arbitrarily large and the system has a finite coherence time.

We give example numbers for trapped ions. Recent experiments have implemented a spin chain with power-law interactions with $J\approx 2\pi\times 1\text{ kHz}$ \cite{richerme14}. We simulate a 1D chain with $\alpha=3$ [Eq.~\eqref{eq:H_1d_power}] and increase $J$ linearly from 0 to $2\pi\times 1\text{ kHz}$ over a time $t_\text{ramp}$. We set $\Omega=2\pi\times 1.04\text{ kHz}$ to maximize the final three-body interaction. The rotating-wave approximation requires $\Omega\gg g$; empirically, we obtain reasonable results with $\Omega=10g$, which corresponds to $g=2\pi\times 104\text{ Hz}.$

\begin{figure}[t]
\centering
\includegraphics[width=4 in,trim=1.6in 4.3in 1in 4.3in,clip]{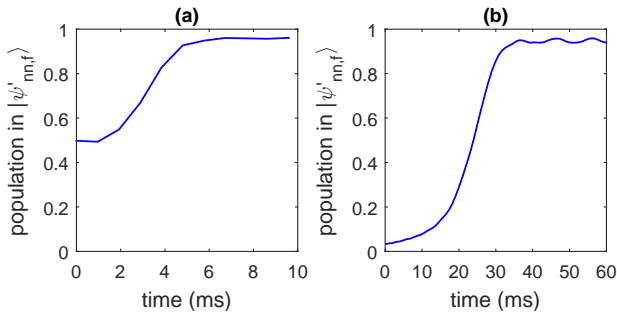}
\caption{\label{fig:adiabatic_experiment}Adiabatic ramp for a 1D chain with power-law interactions ($\alpha=3$) and realistic experimental parameters: (a) $N=3$ with $t_\text{ramp}=7 \text{ ms}$ and (b) $N=8$ with $t_\text{ramp}=40 \text{ ms}$. The wave function is sampled stroboscopically in time at multiples of $t=2\pi/\Omega$. We plot the population in $|\psi'_{nn,f}\rangle$, which is the final ground state of the nearest-neighbor $H'$.}
\end{figure}

The required $t_\text{ramp}$ increases with $N$. For $N=3$, $t_\text{ramp}=4.5/g=7 \text{ ms}$ is sufficient to prepare the ground state of $H'$ with reasonably high fidelity [Fig.~\ref{fig:adiabatic_experiment}(a)]. This is on the order of the coherence time of current experiments \cite{richerme14}. For $N=8$, $t_\text{ramp}=26/g=40 \text{ ms}$ is sufficient [Fig.~\ref{fig:adiabatic_experiment}(b)]. One could decrease $t_\text{ramp}$ by increasing the maximum $J$ or by optimizing the ramp profile.

\section{Modulated $XY$ interactions}

We briefly discuss the effect of time modulation on the $XY$ chain,
\begin{eqnarray}
H&=&J\cos(\Omega t) \sum_n X_n X_{n+1} + g \sum_{n} Y_n Y_{n+1}.\label{eq:H_XY}
\end{eqnarray}
For simplicity, we assume one dimension, nearest-neighbor interactions, and periodic boundary conditions. Such a model can also be implemented with trapped ions, since the $X$ and $Y$ interactions can be independently controlled, although the interaction would be long range \cite{porras04}.

By going into the interaction picture and taking the rotating-wave approximation as in Sec.~\ref{sec:general}, we obtain a time-independent Hamiltonian,
\begin{eqnarray}
H'&=&c_2 \sum_n Y_nY_{n+1} + c_4 \sum_n X_{n-1}Z_nZ_{n+1}X_{n+2}, \label{eq:Hp_XY}\\
c_2 &=& \frac{g}{2}\left[1+\mathcal{J}_0\left(\frac{4J}{\Omega}\right)\right], \label{eq:c2_XY}\\
c_4 &=& \frac{g}{2}\left[1-\mathcal{J}_0\left(\frac{4J}{\Omega}\right)\right]. \label{eq:c4_XY}
\end{eqnarray}
$H'$ contains two- and four-body interactions, where $c_2$ and $c_4$ are exactly the same as $c_1$ and $c_3$ in Eqs.~\eqref{eq:c1_1d} and \eqref{eq:c3_1d}. Equation \eqref{eq:Hp_XY} has been shown to have a second-order phase transition from antiferromagnetic order (${c_4/c_2<1}$) to nematic order ($c_4/c_2>1$) \cite{giampaolo15}, where the latter is also characterized by a nonlocal order parameter.

\section{Conclusion}

We have presented a simple method of generating many-body interactions in atomic systems. One future direction is to consider the effect of disorder. It is known that the transverse-field Ising model with quenched disorder forms a Griffiths phase \cite{fisher92,igloi05}. It would be interesting to see if the Griffiths phase survives time modulation of the interaction. 

Another direction is to add a second slower modulation. For example, suppose one modulated $c_3$ in Eq.~\eqref{eq:Hp_1d} by modulating $J$ on a slower time scale than $1/\Omega$. By going into another rotating frame, one may obtain an even more exotic spin chain.

\begin{acknowledgments}

The simulations in this paper were performed on Indiana University's supercomputer, Big Red II. Y.N.J.~was supported by NSF-DMR Grant No.~1054020.

\end{acknowledgments}

\bibliography{cluster}

\end{document}